\documentclass[11pt]{article}

\usepackage{amssymb,amsfonts}
\usepackage{longtable}
\usepackage{graphicx}

\begin{document}

\title{Mathematical model of immune response to hepatitis B}

\author{F. Fatehi Chenar,\hspace{0.5cm}Y.N. Kyrychko, \hspace{0.5cm}K.B. Blyuss\thanks{Corresponding author. Email: k.blyuss@sussex.ac.uk}\\\\ Department of Mathematics, University of Sussex, Falmer,\\
Brighton, BN1 9QH, United Kingdom}

\maketitle

\begin{abstract}
A new detailed mathematical model for dynamics of immune response to hepatitis B is proposed, which takes into account contributions
from innate and adaptive immune responses, as well as cytokines. Stability analysis of different steady states is performed to identify parameter
regions where the model exhibits clearance of infection, maintenance of a chronic infection, or periodic oscillations. Effects of nucleoside analogues and interferon treatments are analysed, and the critical drug efficiency is determined.
\end{abstract}

\section{Introduction}

Hepatitis B is a major viral infectious disease that affects a third of the world population, with 240-350 million people having a chronic infection \cite{takk09,WHO04}, and over 129 million new infections having occurred since 2013 \cite{lancet386}. This disease is a significant public health burden, causing 750,000 deaths annually \cite{WHO04}, of which about 300,000 can be attributed to liver cirrhosis and hepatocellular carcinoma \cite{lancet385}. Whilst the prevalence of hepatitis B is relatively low (below 1\%) in Western Europe and North America, it remains significant in south-east Asia and sub-Saharan Africa, where 5-10\% of the adult population are chronically infected \cite{WHO04}.

The disease is caused by the hepatitis B virus (HBV), which is a hepatotropic noncytopathic DNA virus of the {\it Hepadnaviridae} family \cite{seeger00}. There are two main routes of transmission of the HBV virus. One is a vertical (perinatal) transmission from an infected mother to a child, resulting in subsequent infection, which in 90\% of cases becomes chronic \cite{liang,reher05}. The other possibility is a horizontal transmission between adults primarily through sexual contacts, intravenous drug use or poor sanitary habits. This type of transmission usually results in recovery, with only 5-10\% of adults developing chronic infections \cite{liang,reher05}. Multiple branches of the immune system are involved in mounting the response during different phases of the HBV infection. In many viral infections of humans, such as HIV, LCMV, Epstein-Barr, the main contribution to the immune response during the early stages of infection comes from the innate immune response, i.e. natural killer (NK) cells and antiviral cytokines, which aim at reducing the spread of the virus and facilitating the development of an adaptive immune response. Contrary to this general observation, early stages of HBV infection are characterised by a delayed viral production and the lack of production of IFN-$\alpha$/$\beta$ \cite{bert06}. Several potential suggestions have been proposed to explain this, including the possibilities that the initial replication of HBV is very slow, or that the virus does not immediately reach the liver and remains for a period of time in other organs \cite{bert06,wie05}, however, the exact mechanism is still largely unknown. Once the exponential phase of HBV expansions properly starts, it activates the innate response and the cytokines \cite{guido99}, which, in turn, induces adaptive immune response, with cytotoxic T lymphocytes (CTLs) being responsible for killing infected cells, and antibodies against HBV surface antigen (HBsAg) neutralizing virus particles and preventing (re)infection of cells. Interestingly, besides killing HBV-infected hepatocytes, CTLs are able to induce non-cytolytic ``cure" of such cells \cite{abbas,guido99,guidotti1}. An important role in the dynamics of immune response against HBV is played by cytokines, which reduce viral replication \cite{devico, isaacs, kalvakolanu}, activate NK and CTL cells \cite{babiker,guidotti1,tamura}, and facilitate induction of immunity in uninfected target cells \cite{ramsay,wiah}.

A number of mathematical models have looked into various aspects of HBV dynamics and that of the immune response during infection. Ciupe et al. \cite{ciupe1,ciupe2} extended a standard model of immune response to study acute HBV infection and the role of time delay associated with activation and expansion of effector cells, and later they also looked into the role of pre-existing or vaccine-induced antibodies in controlling the HBV infection \cite{ciupe3}. Min et al. \cite{min_kuang} have used a standard incidence function rather than a mass action to account for a finite liver size and susceptibility to HBV infection, while Gourley et al. \cite{gourley} have developed a time-delayed extension of this model. Hews et al. \cite{hews} have used a logistic growth for hepatocyte population and a standard incidence to help the model better represent available data and achieve more realistic values for the basic reproduction number. Yousfi et al. \cite{yousfi} have analysed possible mis-coordination between different branches of adaptive immune response, more specifically, the CTLs and the antibodies,  during HBV infection. In terms of the effects of cytokines on mediating immune response, Wiah et al. \cite{wiah} have studied a model that besides the CTLs and antibodies also includes $\alpha$- and $\beta$-interferons, whose role is taken to convert susceptible hepatocytes into infection-resistant cells. Kim et al. \cite{kim12} adapted an earlier model for hepatitis C to include cytokines implicitly through allowing effector cells to cause a non-cytolytic recovery of the infected cells, and a similar approach has also been used by other researchers \cite{dahari,lewin01,sypsa05} who considered a constant rate of non-cytolytic cure alongside treatment.

In this paper we focus on the interplay between various branches of the immune system during HBV infection, with particular emphasis on explicitly modelling the role of cytokines in mediating immune response and controlling viral replication. In the next section we discuss the details of underlying biological processes associated with the immune response against HBV and derive a corresponding mathematical model. Section 3 contains analytical and numerical studies of stability of various steady states. In Section 4 we perform numerical simulations of the model to illustrate different dynamical regimes, as well as to investigate the effects of different types of treatment. The paper concludes with the discussion of results and open questions.    

\section{Model derivation}

\begin{figure}
 \centering
 \includegraphics[scale=0.5]{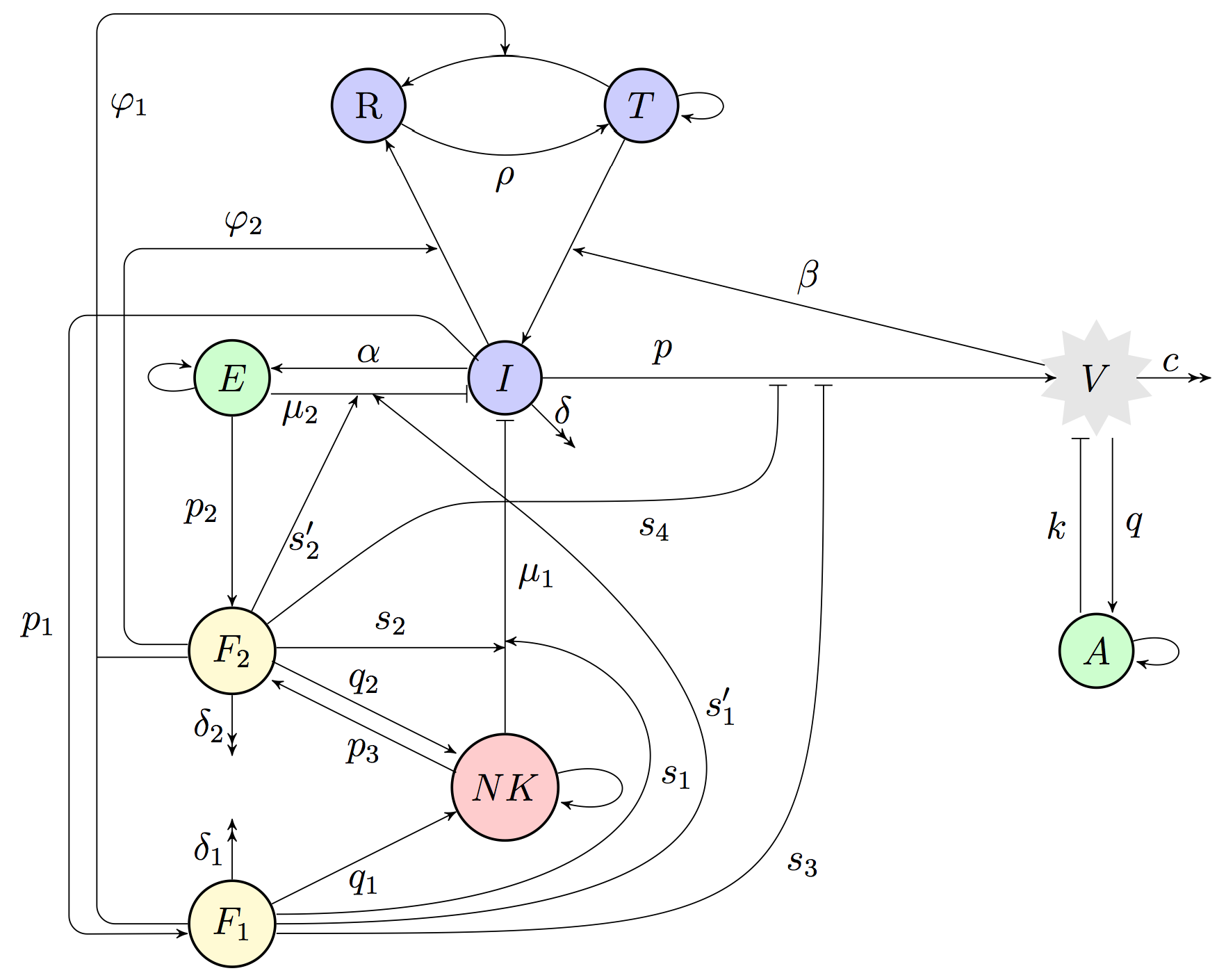}
  \caption{A diagram of immune response to HBV infection. Blue circles indicate host cells (uninfected, infected, and refractory cells), green circles denote adaptive immune response (antibodies, CTLs), yellow circles show cytokines (type-1 and type-2 interferon), red circle is the innate immune response (NK cells), and grey indicates virus particles (virions).
}\label{sys_dia}
\end{figure}

In order to analyse various aspects of immune response to HBV infection, we build on the methodology of some earlier HBV models \cite{nowak,perelson,yongmei}. The host liver cells are divided into populations of uninfected cells $T(t)$, HBV-infected cells $I(t)$, and refractory cells $R(t)$. Healthy hepatocytes are assumed to be produced at a constant rate $\lambda$, die at a rate $d$, and they are infected by virions (free virus particles) at a rate $\beta$. New HBV virions $V(t)$ are produced by the infected cells at a rate $p$, and they are cleared at a rate $c$. Interactions between all cell populations are illustrated in Fig.~\ref{sys_dia}.

Adaptive immune response consists of HBsAg-specific antibodies $A(t)$ that destroy virions at a rate $k$, and HBV-specific CTLs, also referred to as effector cells, $E(t)$. After viral clearance, because of the long-lived plasma and memory B cells, antibody level is kept at some homeostatic level \cite{ciupe3}. To model this, we assume that antibodies are produced at a constant rate $\lambda_a$, and die at per capita rate $d_a$. During infection, antibodies are produced at rate $q$ proportional to the viral load. Whilst antibodies are responsible for eliminating free virus, CTLs instead kill infected cells at a rate $\mu_2$. Some models assume certain basal level of CTLs $s/d_e$ in the absence of infection, where $s$ is the source of CTLs, and $1/d_e$ is their average lifespan \cite{ciupe1, ciupe2}. We will instead assume the dynamics of effector cells in the absence of infection to have the form of logistic growth with the proliferation rate $r_e$ and the carrying capacity $E_{max}$. Upon infection, the immune response is activated, and the population of effector cells will expand at rate $\alpha IE$ \cite{ciupe1, ciupe2}. Similarly to effector cells, in the absence of infection, NK cells are assumed to obey logistic growth with the linear growth rate $r_n$ and the carrying capacity $N_{max}$. 

Let us now focus on the role of cytokines in the immune dynamics. Type-1 interferons IFN-$\alpha/\beta$, to be denoted by $F_1(t)$, are produced by infected cells \cite{busca,guidotti1} at a rate $p_1$, and they are destroyed at a rate $\delta_1$.
Type-2 interferons IFN-$\gamma$, denoted as $F_2(t)$, are produced by CTLs and NKs (natural killer cells) $N(t)$ \cite{guidotti1,devico, guilhot, herbein} at rates $p_2$ and $p_3$, respectively, and they are lost at a rate $\delta_2$. Both types of interferons have the capacity to render the uninfected cells protected from infection through making them resistant to infection \cite{wiah, julkunen,price}, or by turning them into refractory cells \cite{ramsay,ramshaw}. Therefore, the combined effect of interferons making uninfected cells refractory is taken to be $\varphi_1(F_1+F_2)$ per uninfected cell, and refractory cells can lose their viral resistance at a rate $\rho$ \cite{ciupe1}. During infection, IFN-$\alpha/\beta$ are able to activate NK cells \cite{pawelek}, while IFN-$\gamma$ induces protein-10 (CXCL-10) that recruits NK cells \cite{afzal,babiker} and can also activate NK cells \cite{guidotti1}. Hence, the combined effect of interferons on activating NK cells is taken to occur at a rate $q_1NF_1+q_2NF_2$. Besides positive contribution to the production of new NK cells, IFN-$\alpha/ \beta$ also increase the cytotoxicity of NK cells and CTLs \cite{abbas}. On the other hand, IFN-$\gamma$ increases the expression of MHC antigen acting to help CTLs destroy infected cells \cite{tamura}, and it also enhances the activity of NK cells \cite{schroder, carnaud}. Thus, both types of interferons increase cytolytic activity of NKs and CTLs, and hence, we will assume that NKs and CTLs destroy infected cells at rates $\mu_1(1+s_1F_1+s_2F_2)IN$ and $\mu_2(1+s_1^{\prime}F_1+s_2^{\prime}F_2)IE$, respectively. Moreover, antiviral cytokines, such as IFN-$\gamma$ and TNF-$\alpha$, can non-cytopathically purify viruses from infected cells \cite{guidotti1}, so that HBV-specific CTLs and NK cells can effectively ``cure" infected cells through a non-cytolytic antiviral activity mediated by IFN-$\gamma$ \cite{guidotti1, devico, guidotti2, biron}. Hence, infected cells can be lost due to non-cytolytic response of IFN-$\gamma$ at a rate $\varphi_2IF_2$. Studies have shown that IFN-$\gamma$ can activate a number of intracellular mechanisms that suppress viral replication \cite{devico, isaacs, kalvakolanu, stark}, while IFN-$\alpha/\beta$ can stimulate the activation of intracellular antiviral pathways to limit the development and spread of viral replication \cite{guidotti1}. Thus, both types of interferons help infected cells reduce production of new virus particles, so infected cells produce virions at a rate $p/(1+s_3F_1+s_4F_2)$.

With the above assumptions, the complete model for immune response to HBV infection takes the form
\begin{equation}\label{sys1}
\begin{array}{l}
\displaystyle{\frac{dT}{dt}=\lambda-dT-\beta VT+\rho R-\varphi_1T(F_1+F_2),}\\\\
\displaystyle{\frac{dI}{dt}=\beta VT-\delta I-\mu_1(1+s_1F_1+s_2F_2)IN-\mu_2(1+s_1^{\prime}F_1+s_2^{\prime}F_2)IE-\varphi_2IF_2,}\\\\
\displaystyle{\frac{dF_1}{dt}=p_1I-\delta_1F_1,}\\\\
\displaystyle{\frac{dF_2}{dt}=p_2E+p_3N-\delta_2F_2,}\\\\
\displaystyle{\frac{dN}{dt}=r_nN\left(1-\frac{N}{N_{max}}\right)+(q_1F_1+q_2F_2)N,}\\\\
\displaystyle{\frac{dE}{dt}=r_eE\left(1-\frac{E}{E_{max}}\right)+\alpha IE,}\\\\
\displaystyle{\frac{dR}{dt}=\varphi_1T(F_1+F_2)+\varphi_2IF_2-\rho R,}\\\\
\displaystyle{\frac{dV}{dt}=\frac{p}{1+s_3F_1+s_4F_2}I-cV-kAV,}\\\\
\displaystyle{\frac{dA}{dt}=\lambda_a-d_aA-kAV+qV.}\\
\end{array}
\end{equation}

To reduce the complexity of the model and the number of free parameters, we introduce the following rescaled parameters
\[
\begin{array}{l}
\displaystyle{\hat{d}=\frac{d}{r_n},\quad \hat{\beta}=\frac{\beta \lambda_a}{d_ar_n},\quad \hat{\rho}=\frac{\rho \lambda_ad}{r_n\lambda d_a},\quad \hat{\delta}=\frac{\delta}{r_n},\quad \hat{s}_i=s_i\frac{\lambda_a}{d_a},\quad i=1, 2, 3, 4,}\\\\
\displaystyle{\hat{\mu}_1=\frac{\mu_1N_{max}}{r_n},\quad \hat{\mu}_2=\frac{\mu_2E_{max}}{r_n},\quad \hat{\varphi}_i=\frac{\varphi_i\lambda_a}{d_ar_n},\quad \hat{p}_1=\frac{p_1}{r_n},\quad \hat{p}_2=\frac{p_2d_aE_{max}}{r_n\lambda_a},\quad \hat{p}_3=\frac{p_3d_aN_{max}}{r_n\lambda_a},}\\\\
\displaystyle{\hat{r}_e=\frac{r_e}{r_n},\quad \hat{\alpha}=\frac{\alpha \lambda_a}{r_nd_a},\quad \hat{p}=\frac{p}{r_n},\quad \hat{c}=\frac{c}{r_n},\quad \hat{k}=\frac{k\lambda_a}{r_nd_a},\quad \hat{d}_a=\frac{d_a}{r_n},\quad \hat{q}=\frac{q}{r_n},}\\\\
\displaystyle{\hat{s}_i^{\prime}=s_i^{\prime}\frac{\lambda_a}{d_a},\quad \hat{\delta}_i=\frac{\delta_i}{r_n},\quad \hat{q}_i=\frac{q_i\lambda_a}{r_nd_a},\quad i=1,2,}
\end{array}
\]
and new variables
\[
\begin{array}{l}
\displaystyle{\hat{t}=r_nt,\quad T=\frac{\lambda}{d}\hat{T},\quad I=\frac{\lambda_a}{d_a}\hat{I},\quad F_1=\frac{\lambda_a}{d_a}\hat{F}_1,\quad
F_2=\frac{\lambda_a}{d_a}\hat{F}_2,\quad N=N_{max}\hat{N},\quad E=E_{max}\hat{E},}\\\\
\displaystyle{R=\frac{\lambda_a}{d_a}\hat{R},\quad V=\frac{\lambda_a}{d_a}\hat{V},\quad A=\frac{\lambda_a}{d_a}\hat{A}.}
\end{array}
\]
Substituting these variables into the model (\ref{sys1}) and dropping all hats gives the following non-dimensionalised system of equations
\begin{equation}\label{sys2}
\begin{array}{l}
\displaystyle{\frac{dT}{dt}=d(1-T)-\beta V T+ \rho R- \varphi_1 T(F_1+F_2),}\\\\
\displaystyle{\frac{dI}{dt}=\beta VT- \delta I-\left[\mu_1(1+s_1 F_1+ s_2 F_2)N+\mu_2(1+s'_1 F_1+s'_2 F_2)E+\varphi_2F_2\right]I,}\\\\
\displaystyle{\frac{dF_1}{dt}=p_1I-\delta_1F_1,}\\\\
\displaystyle{\frac{dF_2}{dt}=p_2 E+p_3 N-\delta_2 F_2,}\\\\
\displaystyle{\frac{dN}{dt}=N(1-N)+(q_1F_1+q_2F_2)N,}\\\\
\displaystyle{\frac{dE}{dt}=r_e E(1-E)+\alpha IE,}\\\\
\displaystyle{\frac{dR}{dt}=\varphi_1 T(F_1+F_2)+\varphi_2 IF_2-\rho R,}\\\\
\displaystyle{\frac{dV}{dt}=\frac{p}{1+s_3 F_1+s_4 F_2} I-c V-kAV,}\\\\
\displaystyle{\frac{dA}{dt}=d_a(1-A)-kAV+qV.}
\end{array}
\end{equation}

It is straightforward to show that this system is well-posed, i.e. its solutions with non-negative initial conditions remain non-negative for all $t\geq 0$.

\section{Steady states and their stability}

We begin our analysis of the system (\ref{sys2}) by looking at its steady states
\[
S^*=(T^{\ast},I^{\ast},F_1^{\ast},F_2^{\ast},N^{\ast},E^{\ast},R^{\ast},V^{\ast},A^{\ast}),
\]
that can be found by equating the right-hand sides of equations in (\ref{sys2}) to zero and solving the resulting system of algebraic equations. Due to the high dimensionality of the system (\ref{sys2}), it can admit a significant number of possible steady states. Hence, in order to systematically find and analyse all of them, we begin with steady states characterised by the absence of virus particles, i.e. $V^{\ast}=0$, which immediately implies $I^{\ast}=F_1^{\ast}=0$ and $T^{\ast}=A^{\ast}=1$. There are four such steady states,
\[
\begin{array}{l}
\displaystyle{S_1^{\ast}=(1,0,0,0,0,0,0,0,1),\quad S_2^{\ast}=\left(1,0,0,\frac{p_2}{\delta_2},0,1,\frac{\varphi_1p_2}{\rho\delta_2},0,1\right),}\\\\
\displaystyle{S_3^{\ast}=\left(1,0,0,\frac{p_3}{\delta_2-p_3q_2},\frac{\delta_2}{\delta_2-p_3q_2},0,\frac{\varphi_1p_3}{\rho(\delta_2-p_3q_2)},0,1\right),}\\\\
\displaystyle{S_4^{\ast}=\left(1,0,0,\frac{p_2+p_3}{\delta_2-p_3q_2},\frac{\delta_2-p_3q_2+p_2q_2+p_3q_2}{\delta_2-p_3q_2},1,\frac{\varphi_1(p_2+p_3)}{\rho(\delta_2-p_3q_2)},0,1\right).}
\end{array}
\]
Whilst the steady states $S_1^{\ast}$ and $S_2^{\ast}$ are feasible for any values of parameters, $S_3^{\ast}$ and $S_4^{\ast}$ are only biologically feasible, provided $\delta_2-p_3q_2>0$. Linearisation of the system (\ref{sys2}) near each of these steady states shows that $S_1^{\ast}$, $S_2^{\ast}$ and $S_3^{\ast}$ are always unstable, while $S_4^{\ast}$ is stable if the following condition holds
\begin{equation}\label{DF_stab}
K<K_c,\qquad K=\frac{p\beta(\delta_2-p_3q_2)^3}{(c+k)(p_2s_4-p_3q_2+p_3s_4+\delta_2)},
\end{equation}
with
\begin{equation}
\begin{array}{l}
K_c=\delta p_3^2 q_2^2+\mu_1p_2^2q_2s_2-\mu_1p_2p_3q_2^2+\mu_1p_2p_3q_2s_2-\mu_2p_2p_3q_2s'_2+\mu_2p_3^2q_2^2-\mu_2p_3^2q_2s'_2\\\\
-2\delta \delta_2p_3q_2+\delta_2\mu_1p_2q_2+\delta_2\mu_1p_2s_2-\delta_2\mu_1p_3q_2+\delta_2\mu_1p_3s_2+\delta_2\mu_2p_2s'_2-2\delta_2\mu_2p_3q_2\\\\
+\delta_2\mu_2p_3s'_2-p_2p_3q_2\varphi_2-p_3^2q_2\varphi_2+\delta\delta_2^2+\delta_2^2\mu_1+\delta_2^2\mu_2+\delta_2p_2\varphi_2+\delta_2p_3\varphi_2.
\end{array}
\end{equation}
When $K=K_c$, equilibrium $S_4^{\ast}$ undergoes a steady-state bifurcation, and for $K>K_c$, this steady state is unstable.

For $V^{\ast}\neq 0$, one has to distinguish between two cases, $k=q$ and $k\neq q$. For $k=q$, one finds $A^{\ast}=1$, and there are four associated steady states with different combinations of $E^{\ast}=0$ or $E^{\ast}\neq 0$, and $N^{\ast}=0$ or $N^{\ast}\neq 0$. The first of these, $S_5^{\ast}$, characterised by the absence of CTLs and NKs, i.e. $E^{\ast}=0$ and $N^{\ast}=0$, has other components given by
\[
\begin{array}{l}
\displaystyle{T^{\ast}=\frac{(c+k)(dp_1s_3+\delta \delta_1)}{cdp_1s_3+dkp_1s_3+\beta p\delta_1},\quad I^{\ast}=\frac{d\delta_1(p\beta-c\delta-k\delta)}{\delta(cdp_1s_3+dkp_1s_3+\beta p\delta_1)},}\\\\
\displaystyle{F_1^{\ast}=\frac{dp_1(p\beta-c\delta-k\delta)}{\delta(cdp_1s_3+dkp_1s_3+\beta p\delta_1)},\quad F_2^{\ast}=0,}\\\\
\displaystyle{R^{\ast}=\frac{dp_1\varphi_1(c+k)(dp_1s_3+\delta \delta_1)(p\beta-c\delta-k\delta)}{\delta \rho (cdp_1s_3+dkp_1s_3+\beta p\delta_1)^2},
\quad V^{\ast}=\frac{d\delta_1(p\beta-c\delta-k\delta)}{\beta (c+k)(dp_1s_3+\delta \delta_1)},}
\end{array}
\]
and this steady state is always unstable. The steady state $S_6^{\ast}$ with $E^{\ast}=0$ and $N^{\ast}\neq 0$ has components given by
\[
\begin{array}{l}
\displaystyle{I^{\ast}=\frac{\delta_1 F_1^{\ast}}{p_1},\quad F_2^{\ast}=\frac{1+q_1F_1^{\ast}}{a}, \quad N^{\ast}=\frac{\delta_2F_2^{\ast}}{p_3},\quad V^{\ast}=\frac{pI^{\ast}}{(c+k)(1+s_3F_1^{\ast}+s_4F_2^{\ast})},}\\\\
\displaystyle{T^{\ast}=\frac{d+\varphi_2I^{\ast}F_2^{\ast}}{d+\beta V^{\ast}},\quad R^{\ast}=\frac{\varphi_1T^{\ast}(F_1^{\ast}+F_2^{\ast})+\varphi_2I^{\ast}F_2^{\ast}}{\rho},}
\end{array}
\]
where $F_1^{\ast}$ satisfies the cubic equation
\[
b_3(F_1^{\ast})^3+b_2(F_1^{\ast})^2+b_1F_1^{\ast}+b_0=0,
\] 
where the coefficients $b_1$, $b_2$ and $b_3$ are always positive, and
\[
b_0=dp_1\left[-a^3pp_3\beta+(c+k)(a+s_4)(a^2p_3\delta+a\delta_2\mu_1+ap_3\varphi_2+s_2\delta_2\mu_1)\right],\quad a=\frac{\delta_2-p_3q_2}{p_3}.
\]
The steady state $S_6^{\ast}$ is also always unstable.\\

Similarly, the steady state $S_7^{\ast}$ with $E^{\ast}\neq 0$ and $N^{\ast}=0$ has its state variables given by
\[
\begin{array}{l}
\displaystyle{I^{\ast}=\frac{\delta_1 F_1^{\ast}}{p_1}, \quad F_2^{\ast}=\frac{p_2}{\delta_2}\left(1+\frac{\alpha \delta_1F_1^{\ast}}{r_ep_1}\right), \quad E^{\ast}=\frac{r_e+\alpha I^{\ast}}{r_e},\quad V^{\ast}=\frac{pI^{\ast}}{(c+k)(1+s_3F_1^{\ast}+s_4F_2^{\ast})},}\\\\
\displaystyle{T^{\ast}=\frac{d+\varphi_2I^{\ast}F_2^{\ast}}{d+\beta V^{\ast}},\quad R^{\ast}=\frac{\varphi_1T^{\ast}(F_1^{\ast}+F_2^{\ast})+\varphi_2I^{\ast}F_2^{\ast}}{\rho},}
\end{array}
\]
with $F_1^{\ast}$ satisfying the cubic equation
\[
m_3(F_1^{\ast})^3+m_2(F_1^{\ast})^2+m_1F_1^{\ast}+b_0=0,
\]
where $m_1$, $m_2$ and $m_3$ are positive, and
\[
m_0=dp_1^3r_e^3\left[-\beta p\delta_2^2+(c+k)(p_2s_4+\delta_2)(\mu_2p_2s_2^{\prime}+\delta \delta_2+\mu_2\delta_2+\varphi_2p_2)\right].
\]
This steady state is unstable for any parameter values.

The last steady state $S_8^{\ast}$ with $E^{\ast}\neq 0$ and $N^{\ast}\neq 0$ has components
\[
\begin{array}{l}
\displaystyle{I^{\ast}=\frac{\delta_1 F_1^{\ast}}{p_1},\quad E^{\ast}=\frac{r_e+\alpha I^{\ast}}{r_e},\quad F_2^{\ast}=\frac{\alpha p_2\delta_1+r_e p_1\left[
p_2+p_3(1+q)\right]}{r_e p_1 (\delta_2-p_3q_2)},\quad N^{\ast}=\frac{\delta_2F_2^{\ast}-p_2E^{\ast}}{p_3},}\\\\
\displaystyle{V^{\ast}=\frac{pI^{\ast}}{(c+k)(1+s_3F_1^{\ast}+s_4F_2^{\ast})},\quad T^{\ast}=\frac{d+\varphi_2I^{\ast}F_2^{\ast}}{d+\beta V^{\ast}},\quad
R^{\ast}=\frac{\varphi_1T^{\ast}(F_1^{\ast}+F_2^{\ast})+\varphi_2I^{\ast}F_2^{\ast}}{\rho},}
\end{array}
\]
and $F_1^{\ast}$ satisfies a cubic equation. It does not prove possible to determine stability of this steady state in a closed form, so is has to be done numerically.

\newpage

\begin{figure}
	\hspace{0.3cm}
	\includegraphics[scale=0.52]{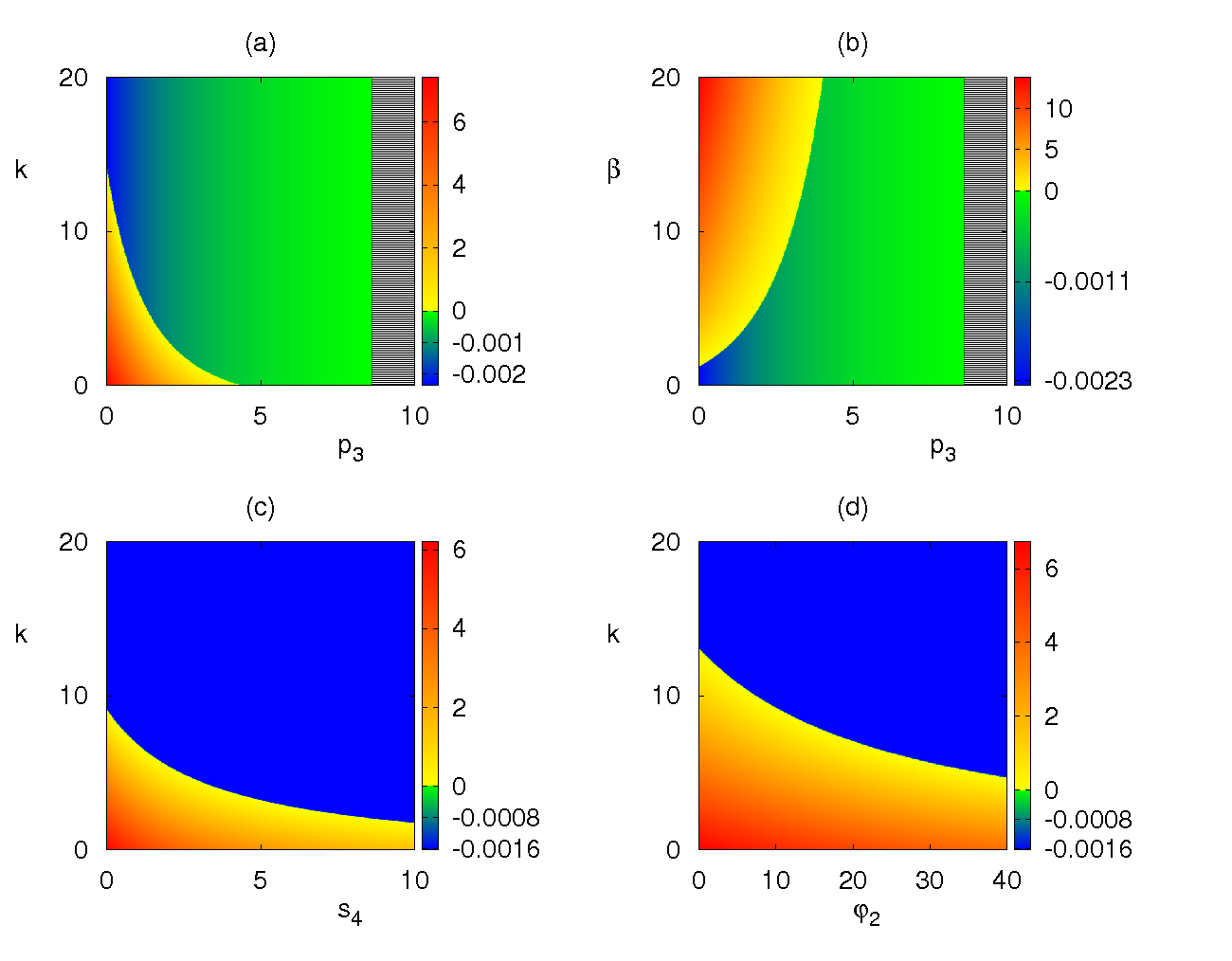}\vspace{-0.5cm}
	\caption{Stability of the disease-free steady state $S_{4}^{\ast}$ with parameter values from Table~\ref{parameter table}. Black grid area indicates the region where there are no feasible steady states. Colour code denotes maximum real part of the largest characteristic eigenvalue for the disease-free steady state $S_4^{\ast}$ when it is feasible.}
	\label{DF}
\end{figure}

For $k\neq q$, we again have four options, depending on whether $E^{\ast}=0$ or $E^{\ast}\neq 0$, and $N^{\ast}=0$ or $N^{\ast}\neq 0$. Similar to the case $k=q$, the steady states $S_9^{\ast}$ with $E^{\ast}=N^{\ast}=0$, $S_{10}^{\ast}$ with $E^{\ast}=0$ and $N^{\ast}\neq 0$ and $S_{11}^{\ast}$ with $E^{\ast}\neq 0$ and $N^{\ast}=0$, are always unstable. The steady state $S_{12}^{\ast}$ with all components being positive cannot be found in a closed form.

The cases $k=q$ and $k\neq q$ have to be considered separately, since for $k\neq q$ one has a relation $V^{\ast}=d_a(1-A^{\ast})/(kA^{\ast}-q)$, which cannot be directly used in the case $k=q$ with $A^{\ast}=1$. However, it is straightforward to show that as $k\to q$, the steady states $S_9^{\ast}$, $S_{10}^{\ast}$, $S_{11}^{\ast}$ and $S_{12}^{\ast}$ converge to $S_5^{\ast}$, $S_6^{\ast}$, $S_7^{\ast}$ and $S_8^{\ast}$, respectively. Of these steady states, only $S_4^{\ast}$ and $S_{12}^{\ast}$ (or equivalently $S_8^{\ast}$ for $k=q$) can potentially change stability, as all other steady states are unstable for any parameter values.

To gain a better understanding of how stability of different steady states is affected by various parameters in the model, we perform numerical stability and bifurcation analysis.
\begin{figure}
	\centering
	\includegraphics[scale=0.58]{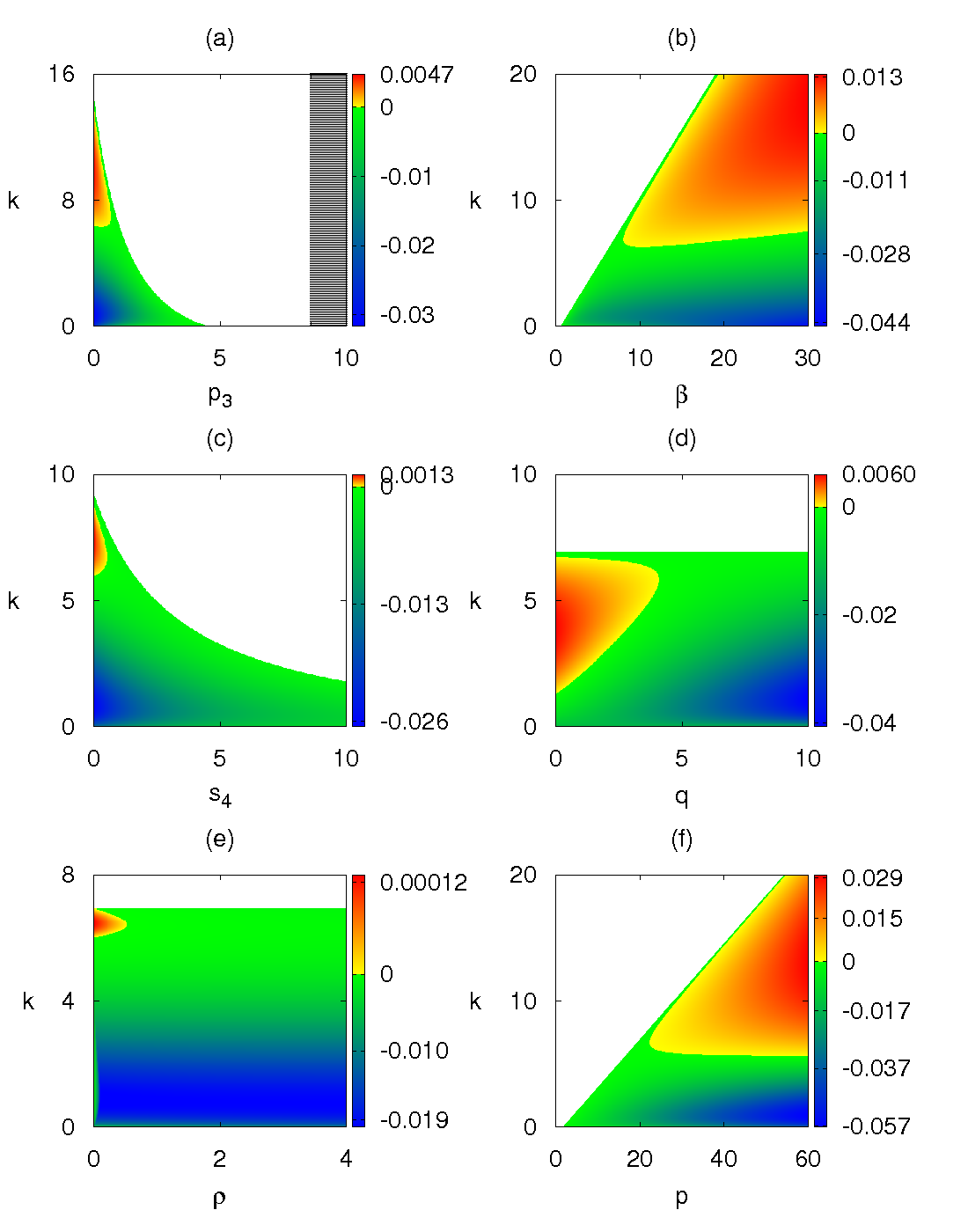}\vspace{-0.5cm}
	\caption{Stability of the endemic steady state $S_{12}^{\ast}$ with parameter values from Table~\ref{parameter table}. White area shows the region where the endemic steady state is not feasible, but the disease-free steady state $S_{4}^{\ast}$ is feasible and stable. Black grid area indicates the region where there are no feasible steady states. Colour code denotes maximum real part of the largest characteristic eigenvalue for the endemic steady state $S_{12}^{\ast}$ when it is feasible.}
	\label{endemic}
\end{figure}
Baseline values of parameters are given in Table~\ref{parameter table} in the Appendix, though one should note that at this stage it is only feasible to explore different qualitative scenarios, as the actual values of many of these parameters have not yet been measured, or significant variations in their values have been reported. 

\begin{figure}
	\begin{center}
	\includegraphics[scale=0.4]{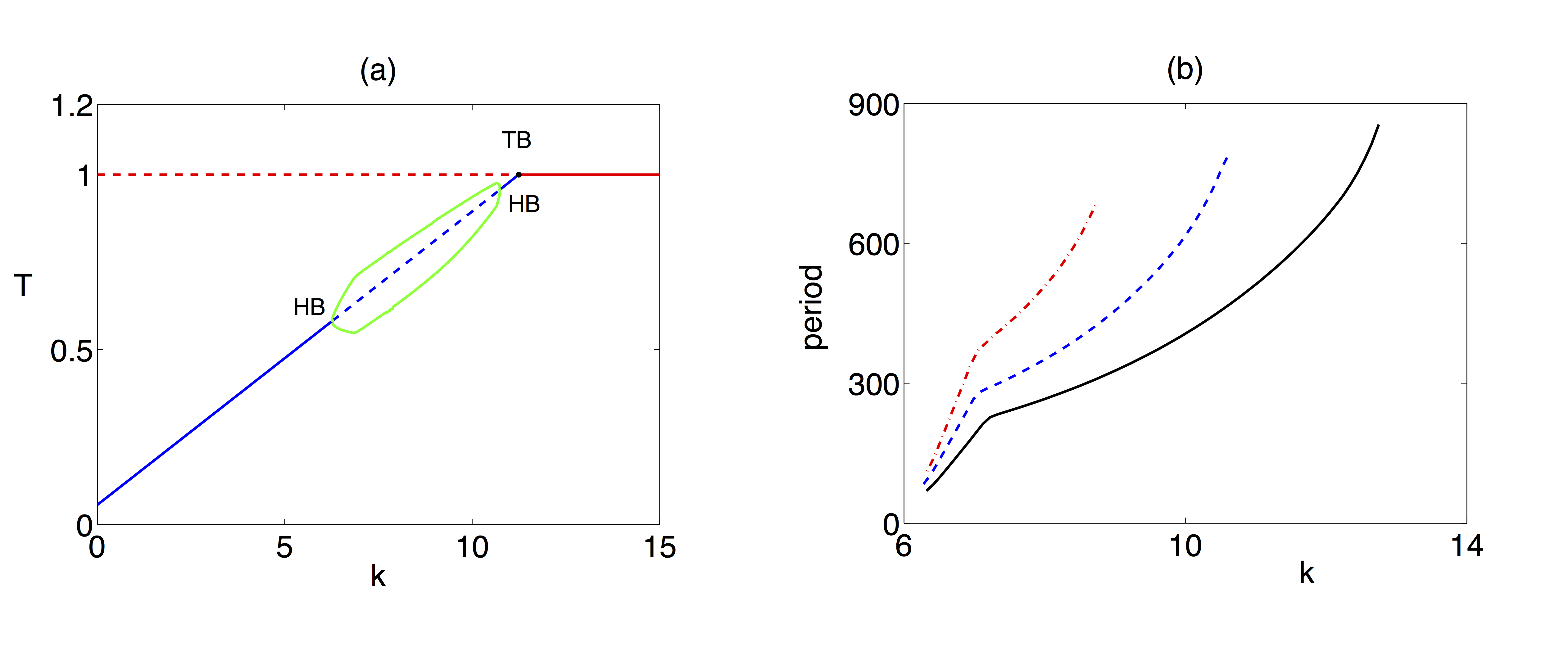}\vspace{-0.7cm}
	\caption{Bifurcation diagram (a) and periods of periodic solutions (b) with parameter values from Table~\ref{parameter table}. (a) In this figure $p_3=0.3$. The blue line shows the endemic steady state, and the red line shows the disease-free steady state, with solid (dashed) lines corresponding to stable (unstable) steady states. At $k=6.277$ and $k=10.74$ there is a Hopf bifurcation of the endemic steady steady state, and at $k=11.2389$ there is a transcritical bifurcation. Between the two HB points there is a stable periodic solution, the minimum and maximum of $T$ are shown in green. (b) This figure shows the dependence of the period of periodic solutions on $k$ for $p_3=0.1$ (black), $p_3=0.3$ (blue), $p_3=0.5$ (red).}
	\label{bifurcation}
	\end{center}
\end{figure}
Figure~\ref{DF} shows regions of feasibility and stability of the disease-free steady state $S_4^{\ast}$. Our earlier analysis indicates that this steady state is only feasible, provided $\delta_2-p_3q_2>0$, which means that this steady state can only exist if the rate $p_3$ of production of IFN-$\gamma$ by NK cells, and the rate $q_2$ at which IFN-$\gamma$ in turn upregulates the production of new NK cells, are not too large, as illustrated in Fig.~\ref{DF}(a) and (b). Stability of the disease-free steady state $S_4^{\ast}$ is determined by the value of $K$ defined in (\ref{DF_stab}), and Figs.~\ref{DF}(a) and (b) suggest that increasing $p_3$ can stabilise this equilibrium if it were previously unstable, which should be expected, as increasing the number of NK cells and the amount of IFN-$\gamma$ leads to a more effective eradication of the viral population. Similarly, increasing the rate of clearance of virions by antibodies $k$, the rate at which IFN-$\gamma$ inhibits production of new virus particles $s_4$, or the rate of IFN-$\gamma$-induced conversion from infected cells to refractory cells $\varphi_2$, all lead to the stabilisation of the disease-free steady state. At the same time, comparison of Fig.~\ref{DF}(a) with (c) and (d) indicates that if antibodies are not very effective, i.e. if $k$ is small, it is easier to clear the infection, i.e. achieve stability of the disease-free steady state, by increasing production of IFN-$\gamma$ by NK cells, since both $s_4$ and $\varphi_2$ have to be increased very significantly before the stability can be achieved.

Figure~\ref{endemic} illustrates how regions of feasibility and stability of the endemic steady state $S_{12}^{\ast}$ depend on system parameters. Comparison of Fig.~\ref{endemic}(a) with Fig.~\ref{DF}(a) suggests that as the disease-free steady state loses its stability, the endemic steady state becomes biologically feasible and stable. However, for very small values of $p_3$, there is a certain range of $k$ values, for which the endemic steady state is also unstable, and one could expect the appearance of periodic solutions.
\begin{figure}
	\centering
	\includegraphics[scale=0.56]{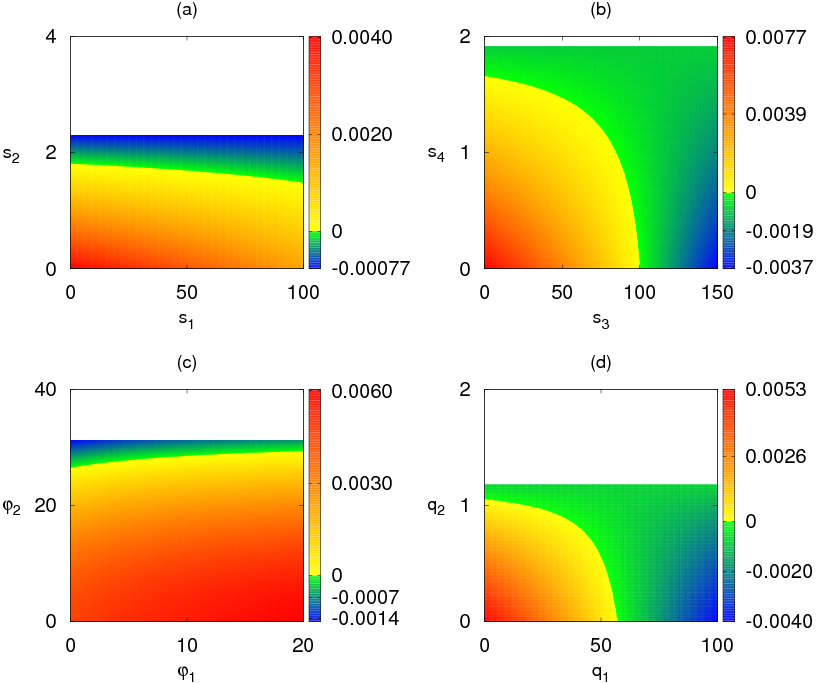}\vspace{-0.2cm}
	\caption{Stability of the endemic steady state $S_{12}^{\ast}$ with parameter values from Table~\ref{parameter table}. White area shows the region where the endemic steady state is not feasible, but the disease-free steady state $S_{4}^{\ast}$ is feasible and stable. Colour code denotes maximum real part of the largest characteristic eigenvalue for the endemic steady state $S_{12}^{\ast}$ when it is feasible.}
	\label{fig_extra}
\end{figure}
This is illustrated in more detail in the bifurcation diagram shown in Fig.~\ref{bifurcation}(a), which indicates that when one fixes some small value of $p_3$ and increases $k$, the endemic steady state does indeed lose its stability via a supercritical Hopf bifurcation, and then regains it at a subcritical Hopf bifurcation for yet higher value of $k$. In the range of $k$ values where the endemic steady state $S_{12}^{\ast}$ is unstable, one observes a stable periodic orbit, whose period increases with $k$ but reduces with $p_3$, as shown in Fig.~\ref{bifurcation}(b). The effects of varying $s_4$ and $\varphi_2$ on stability of $S_{12}^{\ast}$ are similar to those of varying $p_3$, with the exception that for small $k$, increasing $s_4$ or $\varphi_2$ does not make this steady state infeasible, i.e. biologically irrelevant. Figures~\ref{endemic}(b) and (f) are quite similar to each other in that for each value of $k$, there is some minimal value of the infection rate $\beta$ or production rate of new virions $p$, above which the endemic steady state $S_{12}^{\ast}$ becomes biologically feasible and stable. If $k$ is small, then further increases of $\beta$ or $p$ do not have effect on stability, and $S_{12}^{\ast}$ remains stable, whilst for higher $k$ increasing either $\beta$ or $p$ results in the loss of stability through a supercritical Hopf bifurcation. A very interesting behaviour is observed in Fig.~\ref{endemic}(d), which shows that for $k$ small or very large, the stability of $S_{12}^{\ast}$ is unaffected by changes in the rate of production of new antibodies $q$, whereas for an intermediate range of $k$, $S_{12}^{\ast}$ is unstable for small $q$ but gains stability as $q$ is increased. This is quite counter-intuitive, as one would normally expect that if more antibodies are produced for the same viral load, this would help clear the infection. Since $k$ is also the rate at which antibodies are binding free virus and, hence, are removed, this means that it is the balance between $k$ and $q$ that determines whether the infection is maintained at a steady level, i.e. $S_{12}^{\ast}$ is stable, or if periodic oscillations appear in the dynamics. Similar behaviour can be observed in Fig.~\ref{endemic}(e), which shows that the endemic steady state $S_{12}^{\ast}$ is unstable for small $\rho$, i.e. for long periods of viral resistance, but it stabilises as the duration of viral resistance reduces, i.e. for higher values of $\rho$.

In order to better understand the role of cytokines in system's dynamics, we present in Fig.~\ref{fig_extra} stability of the endemic steady state depending on cytokine-related parameters. Figures~\ref{fig_extra}(a) and (b) suggest that increasing the rates $s_1$ and $s_2$ at which IFN-$\alpha/\beta$ and IFN-$\gamma$ enhance cytolytic activity of NK cells, or the rates $s_3$ and $s_4$ at which these interferons inhibit production of new virions, results in stabilisation of the endemic steady state $S_{12}^{\ast}$. One should note, however, that while increasing the rates $s_1$ or $s_3$, associated with IFN-$\alpha/\beta$ only acts to make the endemic steady state more stable, increasing the rates $s_2$ or $s_4$ associated with IFN-$\gamma$ can actually make the endemic steady state biologically irrelevant, thus helping clear the infection by moving the system to a stable disease-free steady state. This suggests the profoundly different effects of IFN-$\alpha/\beta$ and IFN-$\gamma$ on viral dynamics. A similar phenomenon is observed when one investigates the role of cytokines in producing refractory cells from either uninfected or infected cells. Increasing the rate $\varphi_1$ of conversion of uninfected cells into refractory cells, which involves contributions from both types of interferon, results in destabilisation of the endemic steady state. On the other hand, increasing the rate $\varphi_2$ of non-cytolytic cure of infected cells by IFN-$\gamma$ initially stabilises the endemic steady state, but subsequent increase makes the endemic steady state infeasible, thus leading to clearance of infection, as shown in Fig.~\ref{fig_extra}(c). We have also looked into the effects of both types of interferon on enhancing cytotoxic activity of CTLs, as represented by parameters $s'_1$ and $s'_2$. In this case, numerical calculations suggest that the stability of the endemic steady state is not sensitive to $s'_1$, implying that this particular contribution from IFN-$\alpha/\beta$ does not help clear the infection. In this respect, IFN-$\gamma$ plays a more important role, since increasing $s'_2$ above a certain level makes the endemic steady state biologically irrelevant, so the system reverts to a stable disease-free state. Finally, Figure~\ref{fig_extra}(d) shows that increasing the rates $q_1$ and $q_2$ of cytokine-related activation of NK cells leads to stabilisation of the endemic steady state, however, increasing the rate $q_2$ associated with IFN-$\gamma$ beyond certain level results in this steady state becoming biologically irrelevant, thus eradicating the viral infection.

\section{Numerical simulations}

\begin{figure}
	\begin{center}
	\includegraphics[scale=0.6]{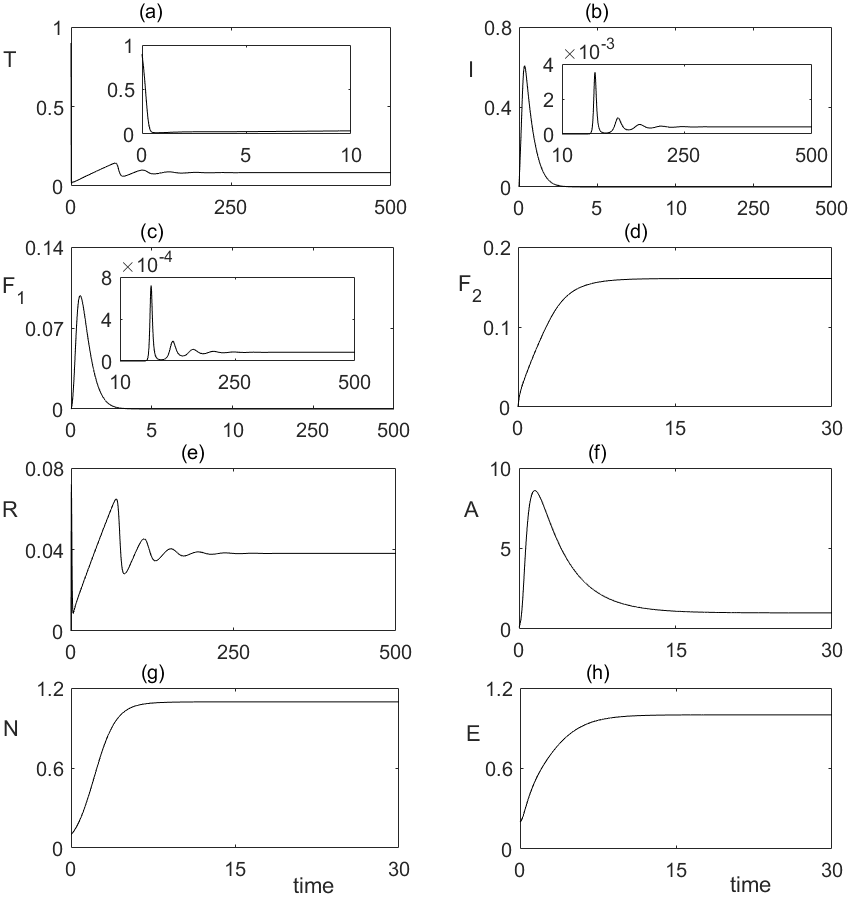}
	\caption{Numerical solution of the model (\ref{sys2}) with parameter values from Table~\ref{parameter table}, and $p_3=0.3$, $k=0.3$. In this case, the system approaches a stable endemic steady state $S_{12}^{\ast}$.}
	\label{NS2}
	\end{center}
\end{figure}

\begin{figure}
	\begin{center}
	\includegraphics[scale=0.55]{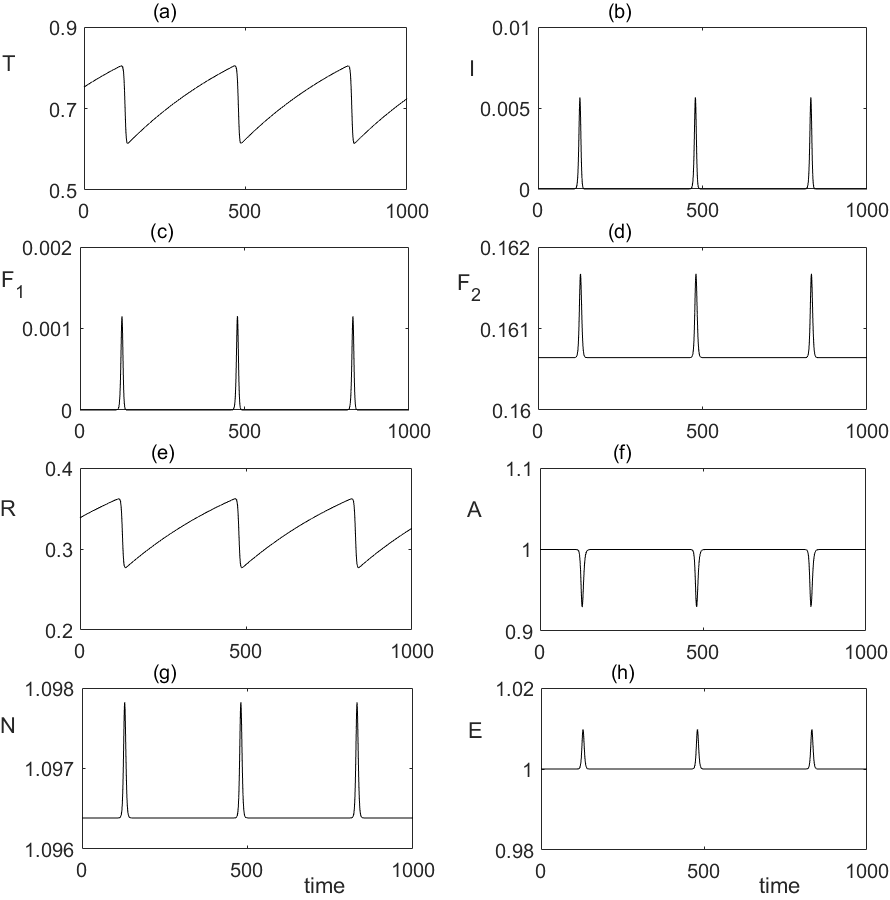}
	\caption{Numerical solution of the model (\ref{sys2}) with parameter values from Table~\ref{parameter table}, and $p_3=0.3$, $k=8$. In this case, both the disease-free $S_{4}^{\ast}$ and the endemic steady state $S_{12}^{\ast}$ are unstable, and the system exhibits a periodic solution.}
	\label{NS3}
	\end{center}
\end{figure}

To demonstrate different types of dynamical behaviour that can be exhibited by the model (\ref{sys2}) in various parameter regimes, we solve this system numerically using the baseline values of parameters given in Table~\ref{parameter table} in the Appendix, and the results are shown in Figs.~\ref{NS1}, \ref{NS2}, \ref{NS3}. In all these figures, the free virus $V(t)$ exhibits the behaviour that is qualitatively similar to that of the number of infected cells, hence, we plot instead the dynamics of the population of refractory cells $R(t)$. Figure~\ref{NS1} illustrates the dynamics of immune response when the condition (\ref{DF_stab}) holds. In this case, the initial viral growth leads to an increase in the numbers of NKs and CTLs, as well as both types of interferons, which results in the successful clearance of the HBV infection, upon which type-1 interferons are also destroyed, and the system settles on a stable disease-free steady state $S_4^*$. Figure~\ref{NS2} shows the dynamics in the case where the endemic steady state $S_{12}^*$ is feasible and stable. One observes that the initial viral growth is suppressed by the combined effects of different branches of the immune system. However, the approach to the endemic steady state is oscillatory with the amplitude of oscillations decaying, with each subsequent viral peak being smaller than the previous one. In the case when the endemic state is unstable due to Hopf bifurcation, one observes stable oscillations, as shown in Fig.~\ref{NS3}. Biologically, these would correspond to the so-called ``flare-ups" \cite{chang14,per01}, where the infection is never completely cleared, but through the interactions between the virus and the immune system, there are periods of very low viral activity followed by the periods of acute viral growth. This situation is reminiscent of the infection-induced autoimmune reaction, where initial viral infection can lead to a breakdown of immune tolerance, so that even in the absence of any exogenous factors or subsequent infections, patients exhibit periods of remission and relapses \cite{BN12,BN15}. It is worth noting that the behaviour shown in Fig.~\ref{NS3} has the hallmarks of slow-fast dynamics, or relaxation oscillations, that are not uncommon in models of immune response \cite{Len00,Mer78}. At every ``flare-up", there is a significant growth in the number of infected cells that triggers the proliferation of both types of interferon, as well as the growth in the populations of CTLs and natural killer cells. All of them are growing very quickly, resulting in a fast immune response that reduces the infection, but as the number of infected cells subsides, so do all the various populations associated with the immune response. Hence, the infection is not completely cleared but rather is kept in check at a very small level. Now, as the population

\begin{figure}
	\begin{center}
	\includegraphics[scale=0.55]{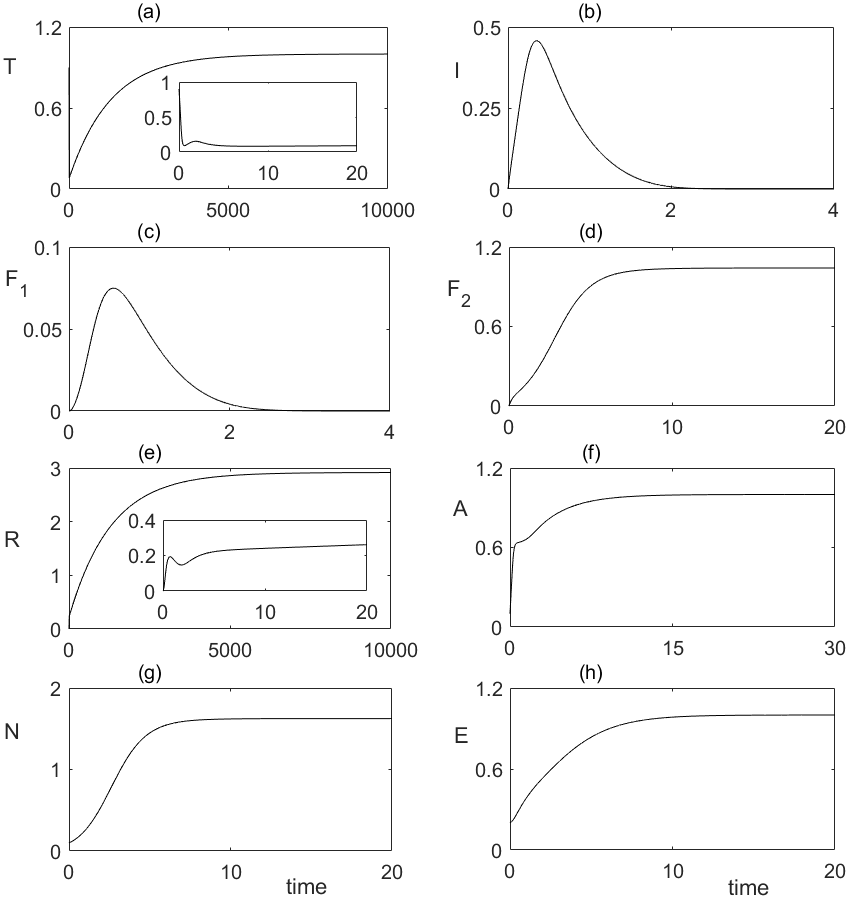}
	\caption{Numerical solution of the model (\ref{sys2}) with parameter values from Table~\ref{parameter table}, and $p_3=3$, $k=8$. In this case, the disease-free steady state $S_4^{\ast}$ is stable, so immune system is able to clear the initial infection.}
	\label{NS1}
	\end{center}
\end{figure}

\noindent of susceptible cells recovers, which is happening on a much longer time-scale, more of these cells become the target of free virus, resulting in a new episode of high viral load, and the cycle repeats.

\begin{figure}
	\centering
	\includegraphics[scale=0.6]{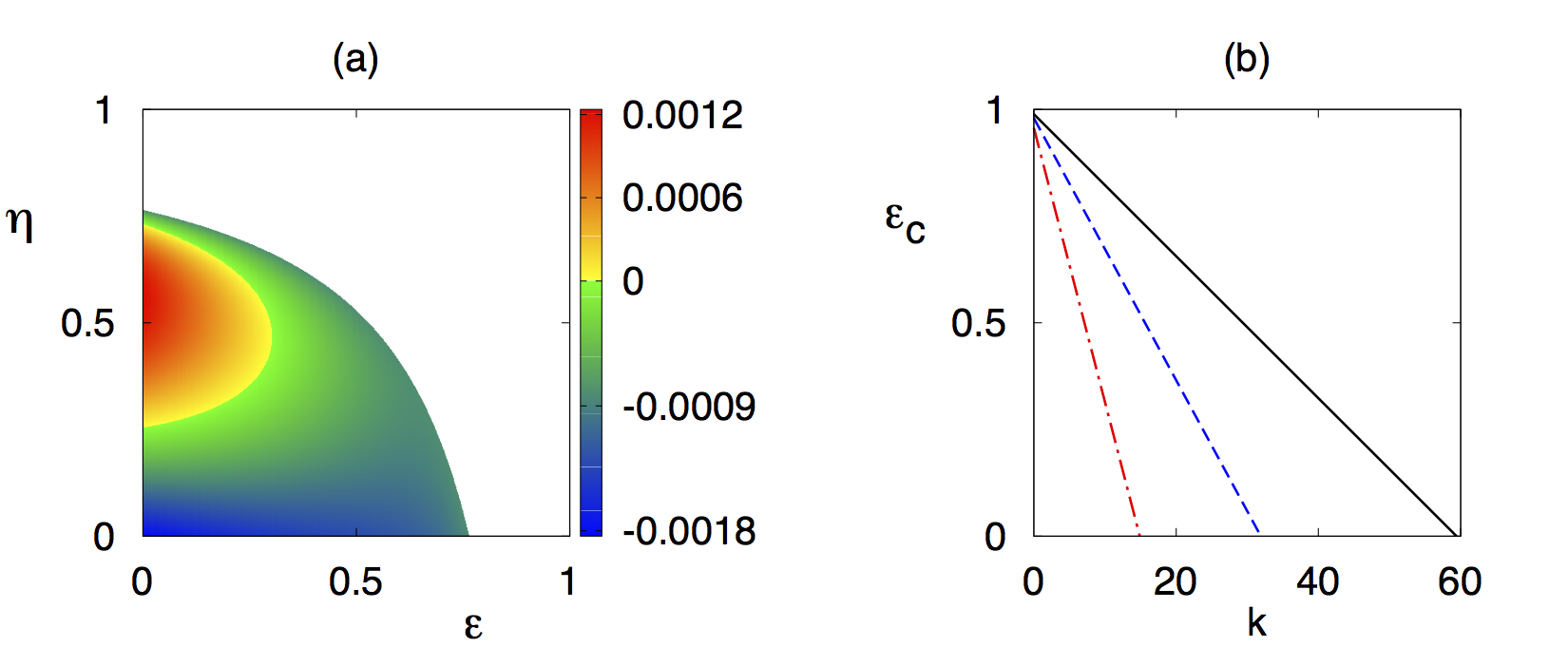}
	\caption{Effects of NAs and interferon therapy on the dynamics of HBV with parameter values from Table~\ref{parameter table}, and $k=7$, $\beta=30$.(a) Stability plot for the endemic steady state $S_{12}^{\ast}$, with the colour code denoting maximum real part of the largest characteristic eigenvalue for the endemic steady state when it is feasible. White area shows the region where the endemic steady state $S_{12}^{\ast}$ is not feasible, and the disease-free steady state $S_{4}^{\ast}$ is stable. (b) Dependence of the critical drug efficacy (${\epsilon}_c$) on $k$, with disease being cleared for $\epsilon_{tot}>\epsilon_c$, with $p_3=0.1$ (black line), $p_3=0.9$ (blue line), $p_3=2$ (red line).}
	\label{treatment}
\end{figure}

As a next step, we look into effects of antiviral treatments on HBV. There are two main types of drugs used to treat HBV infection: nucleot(s)ide analogues (NAs), such as lamivudine, adefovir, entecavir, tenofovir, telbivudine, famciclovir, telbivudine, clevudine, and IFN-based therapy, which includes stand-alone IFN-$\alpha$ (roferon, intron) or pegylated interferon peg-IFN-$\alpha$2a/2b \cite{dahari,kim12,packer,sypsa05,takk09}. These treatments individually \cite{min,nowak} and in combinations \cite{colo06,lewin01} result in either reducing the production of new virus particles, or in blocking {\it de novo} infections. Mathematically, one can represent these two effects by a modified viral production rate $(1-\epsilon)p$ and a modified transmission rate $(1-\eta)\beta$, where $0\leq\epsilon\leq 1$ and $0\leq\eta\leq 1$ are drug efficacies associated with inhibiting viral production and preventing new infections, respectively. In order to characterise the overall effectiveness of treatment, it can be helpful to consider a cumulative parameter describing the total drug effectiveness ${\epsilon}_{tot}$, which is defined as $1-\epsilon_{tot}=(1-\eta)(1-\epsilon)$ \cite{dahari}. This would allow one to determine a critical drug efficacy, $\epsilon_c$, corresponding to stability boundary of the disease-free steady state $S_4^{\ast}$, so that this steady state would be stable for $\epsilon_{tot}>\epsilon_c$. With these modifications, new equations for the numbers of healthy and infected cells, as well as the free virus, have the form
\begin{equation}\label{treat}
\begin{array}{l}
\displaystyle{\frac{d{T}}{d{t}}={d}(1-{T})-\beta (1-\eta) VT+{\rho} {R}-{\varphi}_1 {T}({F}_1+{F}_2),}\\\\
\displaystyle{\frac{d{I}}{d{t}}={\beta}(1-\eta){V}{T}-{\delta} {I}-{\mu}_1(1+{s}_1{F}_1+{s}_2{F}_2){I}{N}-{\mu}_2(1+{s}_1^{\prime}{F}_1+{s}_2^{\prime}{F}_2){I}{E}-{\varphi}_2{I}{F}_2,}\\\\
\displaystyle{\frac{d{V}}{d{t}}=\frac{{p}(1-\epsilon)}{1+{s}_3{F}_1+{s}_4{F}_2}{I}-{c}{V}-{k}{A}{V},}
\end{array}
\end{equation}
with the rest of the equations remaining the same as in the main model (\ref{sys2}).

\begin{figure}
	\begin{center}
	\includegraphics[scale=0.65]{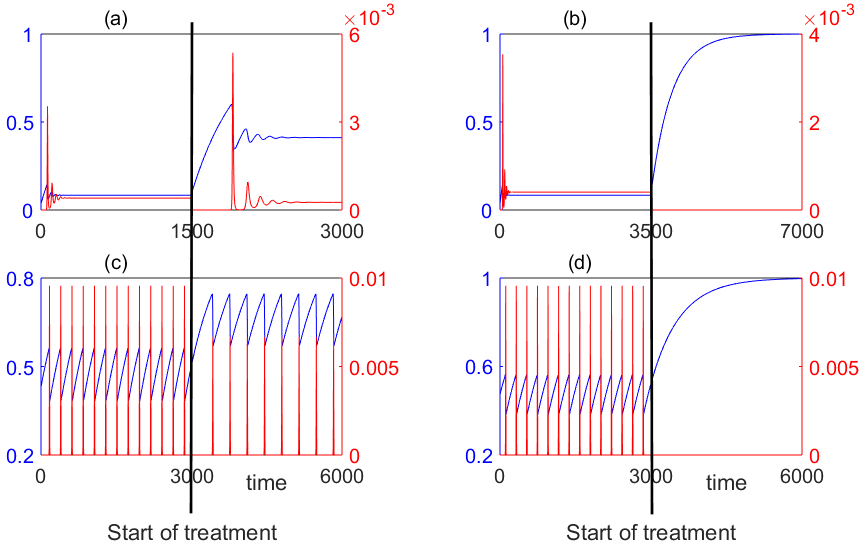}
	\caption{Numerical solution of the model (\ref{sys2}) with treatment (\ref{treat}) and parameter values from Table~\ref{parameter table} and $(k=0.3$, $p=0.3)$ in (a-b), and $(k=8$, $p=0.3)$ in (c-d). In all plots, blue colour denotes a rescaled number of uninfected cells $T(t)$, and red colour denotes a rescaled number of infected cells $I(t)$. (a)-(b) Treatment of the chronic infection with ${\epsilon}_{tot}<\epsilon_c$ $(\eta=0.6$, $\epsilon=0.5)$ (a), and ${\epsilon}_{tot}>\epsilon_c$ $(\eta=0.9$, $\epsilon=0.6)$ (b). (c)-(d) Treatment of the relapsing infection with ${\epsilon}_{tot}<\epsilon_c$ $(\eta=0.2$, $\epsilon=0.1)$ (c), and ${\epsilon}_{tot}>\epsilon_c$ $(\eta=0.2$, $\epsilon=0.4)$ (d).}
	\label{num_sim_treat}
	\end{center}
\end{figure}

Figure \ref{treatment} (a) shows that for parameter values from Table~\ref{parameter table}, if $\eta > 0.7646$, then pure NAs therapy is sufficient to destabilise the endemic steady state and thus clear the infection, and similarly, if $\epsilon > 0.7646$, then just IFN-therapy can make the disease-free steady state $S_4^{\ast}$ stable. This Figure also suggests that disease clearance can be achieved if the combined efficacy ${\epsilon}_{tot}$ exceeds some critical value $\epsilon_c$. Figure \ref{treatment} (b) illustrates how this critical combined efficacy $\epsilon_c$ varies with the rate $k$ of clearance of free virus by antibodies and the rate $p_3$ of production of type-2 interferons by NK cells. One observes that the critical combined efficacy $\epsilon_c$ decreases with $k$, implying that the faster the free virus is cleared by antibodies, the less stringent is the requirement on the efficacy of treatment to clear the infection, and for sufficiently high $k$ the disease clearance can be achieved even in the absence of treatment. Surprisingly, for the same value of $k$, having a higher rate of production of type-2 interferons by NK cells requires a higher combined efficacy $\epsilon_c$ for viral clearance.

Figure~\ref{num_sim_treat} illustrates the effect of using combined NAs and interferon therapy on chronic and relapsing HBV infections. In both regimes, application of treatment with sub-optimal efficacy, i.e. with ${\epsilon}_{tot}<\epsilon_c$, does not cause qualitative change in the system dynamics but results in an increased number of uninfected cells and a decreased number of infected cells. On the contrary, for ${\epsilon}_{tot}>\epsilon_c$, in both cases the number of infected cells is reduced to zero, and the system approaches a stable disease-free steady state $S_4^{\ast}$, which corresponds to a successful clearance of infection. 

\section{Discussion}

In this paper we have derived and analysed a new model for HBV infection with particular emphasis on interactions between different branches of immune system, including innate immune response as exemplified by NK cells, adaptive immune response represented by HBV-specific cytotoxic T cells and antibodies, and various cytokines. During infection the cytokines play an important role in recruitment of innate and adaptive immune factors, and they also help them to be more effective, as well as facilitate non-cytolytic cure of infected cells.

Stability analysis of the steady states has shown how various parameters affect the dynamics of immune response, with some of the results being intuitively clear, and others being quite unexpected. Naturally, increasing the number of NK cells, the rate of clearance of free virus by antibodies, the rate of inhibition of viral production by IFN-$\gamma$, or the rate of conversion from infected to refractory cells, all facilitate a more efficient clearance of infection, making the disease-free steady state stable. Once the disease-free steady state loses its stability, the endemic equilibrium becomes biologically feasible and stable. For sufficiently small values of the rate of production of IFN-$\gamma$ by NK cells, the endemic steady state can lose its stability via Hopf bifurcation, giving rise to stable periodic solutions. We have found that for a very small or a very large rate of free virus clearance by antibodies, the stability of the endemic steady state is unaffected by how quickly the new antibodies are produced, whereas for an intermediate range of virus clearance rate, this steady state is unstable for low production of antibodies, and gains stability as the rate of antibody production is increased. This is a very surprising result, as normally one would expect that a higher rate of production of antibodies for the same viral load leads to a clearance of infection, rather than stabilisation of a chronic state. The implication of this observation is that it is not the individual rates of production of antibodies and viral clearance, but rather the balance between them that determines whether the system maintains a chronic infection or exhibits periodic oscillations.

In terms of the role of cytokines on mediating various branches of immune response, a surprising result of the analysis is that increasing the rates at which IFN-$\alpha/\beta$ and IFN-$\gamma$ increase cytolytic activity of NK cells or inhibit production of free virus, actually leads to stabilisation of the endemic steady state. The major difference in the effects of cytokines IFN-$\alpha/\beta$ and IFN-$\gamma$ lies in the observation that whilst increasing the rates associated with IFN-$\alpha/\beta$ just results in the stabilisation of an otherwise unstable endemic steady state, increasing the same rates for IFN-$\gamma$ can result in making the endemic steady state biologically irrelevant, thus qualitatively changing the dynamics. The same result holds for IFN-$\gamma$-facilitated non-cytolytic cure of infected cells. If the production of IFN-$\gamma$ by NK cells is too high, this makes all steady states of the system unstable, leading to persistent oscillations, thus maintaining the infection.

We have also looked into modelling the dynamics of HBV treatment with nucleot(s)ide analogues and/or stand-alone or pegylated interferons. Since these treatments are known to act by reducing the appearance of new infections and blocking production of free virus, we have looked at how the combined drug efficacy depends on these two properties. Numerical studies have shown the existence of a minimum drug efficacy required to clear the infection, and, unexpectedly, this critical drug efficacy is actually increasing with the rate of production of IFN-$\gamma$ by NK cells.

There are several directions in which the model presented in this paper can be extended. One important aspect of the immune dynamics is the non-instantaneous nature of several important processes, such as the lag between infection and recruitment of CTLs, production of new virus particles once a cell becomes infected, the time required for viral cell entry etc \cite{beau08,nelson00}. Mathematically, this can be represented by including discrete of distributed time delay for each of the associated processes, which would make the model more realistic but would also make the analysis much more involved. Furthermore, it is known that antibodies do not kill the virus particles directly, but rather stick to them, creating a virus-antibody complex \cite{ciupe3}. These complexes are not stable forever and can experience some dissociation, hence, explicitly including them into the model can provide better insights into the dynamics.

\section*{Acknowledgements.} FFC acknowledges the support from Chancellor's Studentship from the University of Sussex.

\bibliographystyle{ieeetr}
\bibliography{sample}

\section*{Appendix} The parameter values used for numerical simulations are given in the table below.\\

\begin{table}
		\caption{Table of baseline parameter values.}
		\label{parameter table}

\begin{center}
\begin{tabular}{|l|l|l|}
\hline
Parameter & Value & Definition \\
\hline
	$d$ & 0.003 & Natural death rate of uninfected cells \\
	\hline
	$\beta$ & 7 & Infection rate \\
	\hline
	$\rho$ & 5 & Rate of missing refractory state \\
	\hline
	$\varphi_1$ & 14 & Rate of IFN-induced conversion from uninfected cells to refractory cells \\
	\hline
	$\delta$ & 0.56 & Natural death rate of infected cells \\
	\hline
	$\mu_1$ & 5 & Death rate of infected cells by NK cells \\
	\hline
	$s_1$ & 1.5 & Effect of IFN-$\alpha/\beta$ on NK cells to kill infected cells \\
	\hline
	$s_2$ & 0.6 & Effect of IFN-$\gamma$ on NK cells to kill infected cells \\
	\hline
	$\mu_2$ & 0.14 & Death rate of infected cells by HBV-specific CTLs \\
	\hline
	$s^{\prime}_1$ & 1.9 & Effect of IFN-$\alpha/\beta$ on the HBV-specific CTLs\\
	\hline
	$s^{\prime}_2$ & 2 & Effect of IFN-$\gamma$ on the HBV-specific CTLs\\
	\hline
	$\varphi_2$ & 21 & Rate of IFN-$\gamma$-induced conversion from infected cells to refractory cells \\
	\hline
	$p_1$ & 1 & Production rate of IFN-$\alpha/\beta$ by infected cells \\
	\hline
	$\delta_1$ & 4.9 & Natural death rate of IFN-$\alpha/\beta$ \\
	\hline
	$p_2$ & 0.5 & Production rate of IFN-$\gamma$ by HBV-specific CTLs \\
	\hline
	$p_3$ & 0.9 & Production rate of IFN-$\gamma$ by NK cells \\
	\hline
	$\delta_2$ & 5.16 & Natural death rate of IFN-$\gamma$ \\
	\hline
	$q_1$ & 0.8 & Production rate of NK cells by IFN-$\alpha/\beta$ \\
	\hline
	$q_2$ & 0.6 & Production rate of NK cells by IFN-$\gamma$ \\
	\hline
	$r_e$ & 0.5 & Maximal growth rate of HBV specific cytotoxic T cells \\
	\hline
	$\alpha$ & 1 & Antigen-dependent proliferation rate of HBV-specific CTLs \\
	\hline
	$p$ & 20 & Production rate of free virus \\	
	\hline
	$s_3$ & 1.7 & Effect of IFN-$\alpha/\beta$ on the production of free viruses \\
	\hline
	$s_4$ & 1 & Effect of IFN-$\gamma$ on the production of free viruses \\
	\hline
	$c$ & 0.67 & Natural clearance rate of free viruses \\
	\hline
	$k$ & 2 & Clearance rate of free viruses by antibodies\\
	\hline
	$d_a$ & 0.332 & Natural death rate of free antibodies \\
	\hline
	$q$ & 5 & Production rate of free antibody by free viruses \\
	\hline
\end{tabular}
\end{center}
\end{table}

\end{document}